\documentclass[
aps,%
12pt,%
final,%
notitlepage,%
oneside,%
onecolumn,%
nobibnotes,%
nofootinbib,%
superscriptaddress,%
noshowpacs]%
{revtex4}
\usepackage{color}

\usepackage{soul}
\usepackage{graphicx,rotating,amsmath,amsfonts,amssymb}

\usepackage{natbib}
\usepackage[utf8]{inputenc}

\newcommand{\Mpc}{$h^{-1}$\thinspace Mpc}

\newcommand{\be}{\begin{equation}}
\newcommand{\ee}{\end{equation}}
\newcommand{\dd}[1]{{\rm d}#1\,}

\newcommand{\ba}{\begin{array}} 
\newcommand{\ea}{\end{array}}

\def\apj{ApJ}
\def\aj{AJ}
\def\apjl{ApJL}

\def\aap{A\&A}

\def\mnras{MNRAS}

\def\nat{Nature}

\begin{document}

\noindent {\it Astronomy Reports, 2026, Vol. , No. }
\bigskip\bigskip  \hrule\smallskip\hrule
\vspace{35mm}

\title{FRACTAL PROPERTIES OF THE COSMIC WEB\footnote{Paper presented
    at the Sixth Zeldovich 
    meeting, an international conference in honor of Ya. B. Zeldovich
    held in Pescara, Italy on July 13--17, 2026. Published by the
    recommendation of the special editors: R. Ruffini and
    G. V. Vereshchagin.}}

\author{\bf \copyright $\:$  2026.
\quad \firstname{Jaan}~\surname{Einasto,}}%
\email{jaan.einasto@ut.ee}
\affiliation{Tartu Observatory, University of Tartu, Estonia}%

\begin{abstract}

\centerline{\footnotesize Received: ;$\;$
Revised: ;$\;$ Accepted: .}\bigskip\bigskip\bigskip

The cosmic web is one of the most complex systems in nature,
consisting of galaxies and clusters of galaxies connected by filaments
and walls, and separated by large empty regions known as cosmic voids.
The most common method for describing the web is the correlation
function and its derivative, the structure function and fractal
dimension function. In this paper I
review the fractal properties of the cosmic web within the concordance
$\Lambda$CDM framework. I describe how the fractal function is derived
from the angular and spatial distributions of galaxies and discuss the
relations between these approaches.

\end{abstract}

\maketitle

\section{Introduction \label{intro}}

Astronomers in Tartu had a long and fruitful collaboration with Yakov
Borisovich Zeldovich. To advance physical cosmology, Zeldovich
organized regular summer and winter schools. One of the earliest
summer schools was held at the new Tartu Observatory in Tõravere,
Estonia, in the summer of 1962. Later, similar schools took place in
Caucasus vacation homes. He also led bi-weekly seminars at the
Sternberg Astronomical Institute in Moscow.

To stimulate international discussion on cosmology, Zeldovich proposed
organizing international conferences, suggesting Tallinn, Estonia, as a
venue. This initiative led to the IAU Symposium on ``Large Scale
Structure of the Universe'' \citep{Longair:1978}. At this symposium,
the existence of the cosmic web was independently identified by four
research groups. The next IAU Symposium, held in Greece in 1982,
focused on quantitative tests of theoretical models for the formation
of the web. The last cosmology symposium Zeldovich attended before his
unexpected passing was held in Balatonfured, Hungary, in the summer of
1987, where the fractal nature of the Universe was discussed for the
first time.

In this review I discuss the fractal properties of the cosmic web in
the observed  Universe, following \citet{Einasto:2025ac}. I adopt the
standard $\Lambda$CDM cosmology with parameters
($\Omega_m,\Omega_{\Lambda},\Omega_b,h,\sigma_8,n_s$)
=(0.28,~0.72,~0.044,~0.693,~0.84,~1.00).

\section{A short history of fractal studies of the Universe \label{hist}}

Scientists have long recognized that many natural processes exhibit
self-similarity across a wide range of scales. Classic examples include
coastlines and mountain landscapes. Benoit \citet{Mandelbrot:1977aa}
introduced the term {\em fractal} to describe such phenomena.

Self-similarity in the distribution of galaxies and other astronomical
objects was noted already by \citet{Charlier:1922aa} and later studied
in more detail by \citet{Carpenter:1938aa}, \citet{Kiang:1967},
\citet{Haggerty:1972aa}, and \citet{de-Vaucouleurs:1970}.

\subsection{Angular distribution of galaxies \label{angular}}

\begin{figure}[h]
\centering
\resizebox{0.48\textwidth}{!}{\includegraphics*{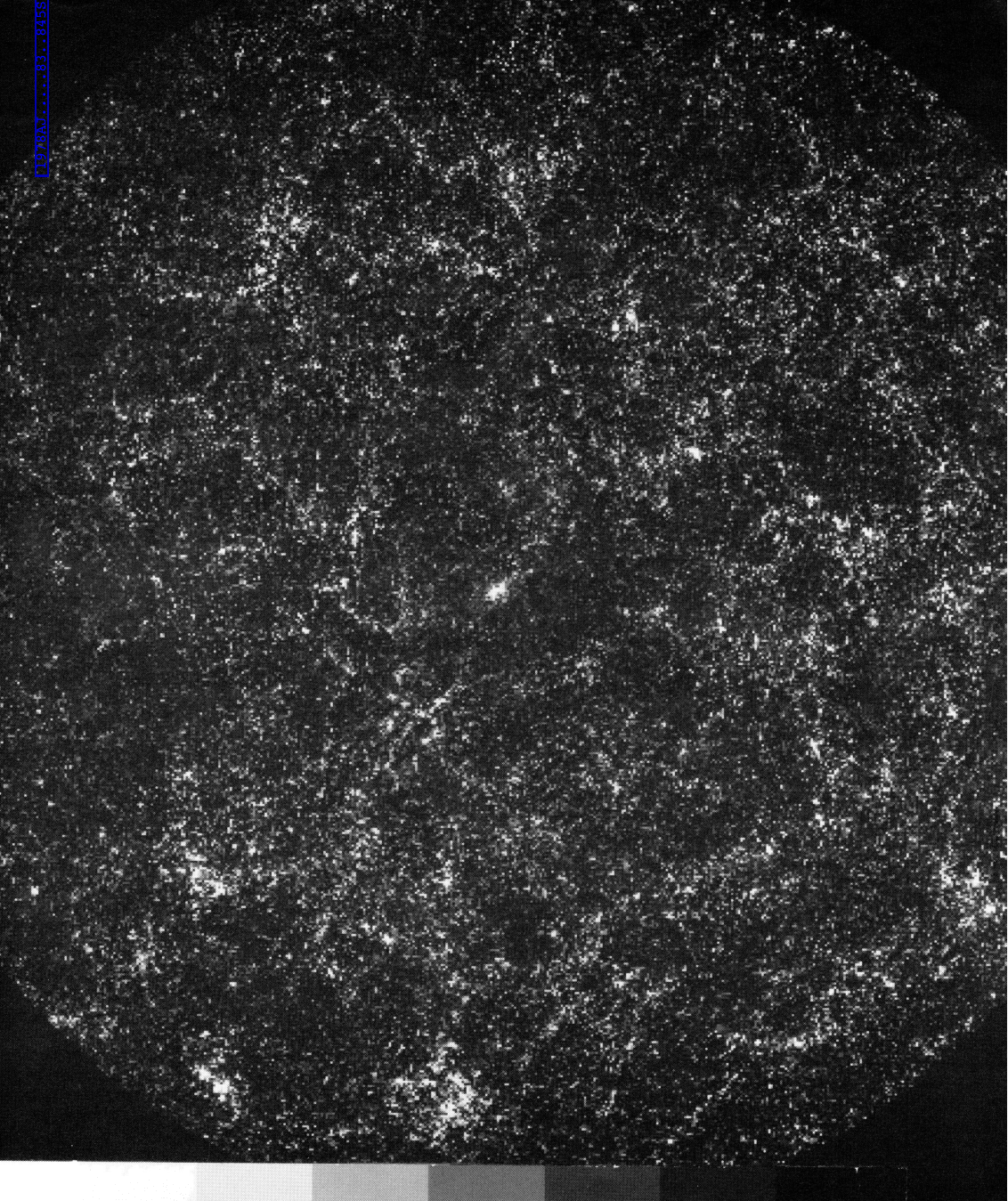}}
\resizebox{0.46\textwidth}{!}{\includegraphics*{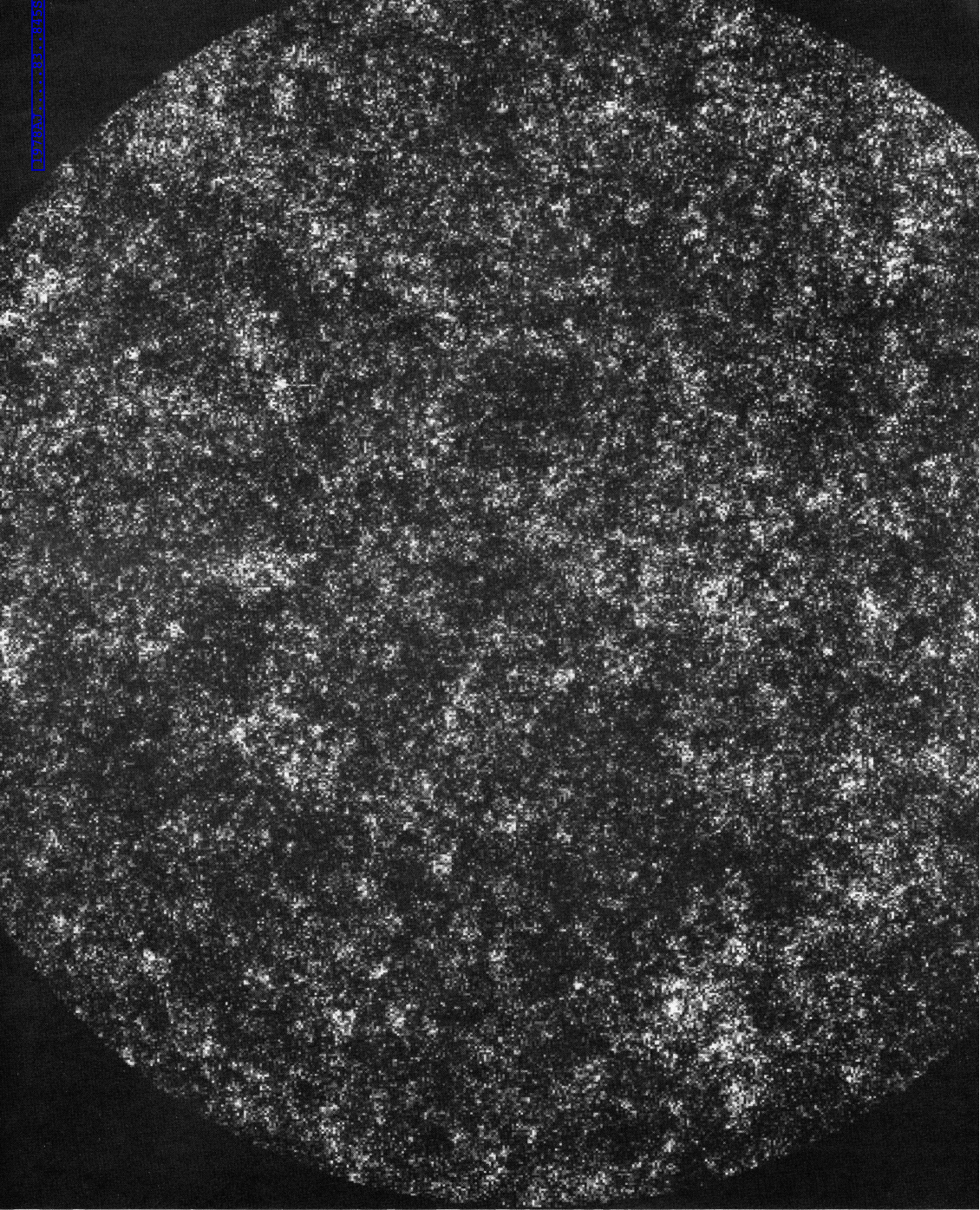}}
\caption{Left: Map of Lick survey galaxies in the northern galactic hemisphere brighter
  than $m_B\le 18.9$ and north of galactic latitude $b\ge40^\circ$.
  Right: Simulated map of galaxies imitating the 2D distribution of
  Lick galaxies. Credit: Jim Peebles  (\citet{Soneira:1978fk}).}
\label{fig:Lick}
\end{figure}

The first deep catalogue of galaxies covering the northern hemisphere
was compiled at the Lick Observatory using the 20-inch Carnegie
astrograph \citep{Shane:1967}. Galaxy counts were made in $10'\times
10'$ cells. \citet{Seldner:1977a} corrected these counts for plate
sensitivity variations and other observational effects. The resulting
map is shown in the left panel of Fig.~\ref{fig:Lick}.

\citet{Soneira:1978fk} constructed a fractal model Universe to match
the observed angular distribution of Lick galaxies. The model places
“galaxies’’ in a hierarchical three-dimensional clustering pattern,
assigns absolute magnitudes, and projects objects brighter than
$m=18.9$ onto the sky. The resulting simulated map is shown in the
right panel of Fig.~\ref{fig:Lick}. Both maps were used to compute
two-point angular correlation functions (CFs).

\begin{figure}[h]
\centering
\resizebox{0.55\textwidth}{!}{\includegraphics*{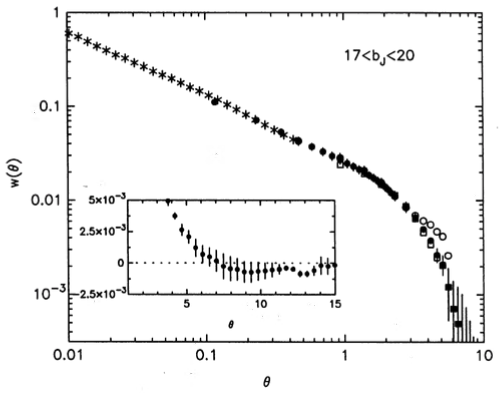}}
\caption{Average angular CF of APM galaxies in the magnitude range
  $17 \le b_j \le 20$. The inset shows the mean angular CF on a linear
  scale. Angular separation is given in degrees   (\citet{Maddox:1990aa}).}
\label{MaddoxFig}
\end{figure}

In the 1970s and 1980s, British and Australian astronomers used
Schmidt telescope plates to photograph the entire sky. The Automatic
Plate Measuring (APM) machine in Cambridge scanned these plates, and
specialized software separated stars from galaxies. The final
catalogue contains over two million galaxies brighter than $b_j
=20.5$.  \citet{Maddox:1990aa} used these data to compute the angular
CF shown in Fig.~\ref{MaddoxFig}.

These studies demonstrated that the angular distribution of galaxies
is well described by a power-law CF,
\be
\xi(r)=(r/r_0)^{-\gamma},
\label{xilaw}
\ee
with correlation length $r_0 =4.5 \pm 0.5$~\Mpc\ and slope
$\gamma = 1.77$. This relation holds over the interval
$0.05 \le r \le 10$~\Mpc\ \citep{Groth:1977aa}.

\subsection{Discussion of the fractal character of the cosmic web}

\begin{figure}[h]
\centering
\resizebox{0.50\textwidth}{!}{\includegraphics*{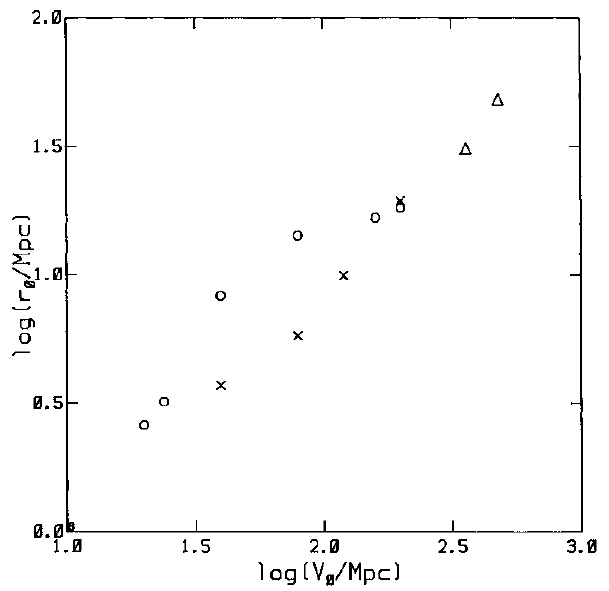}}
\caption{Correlation lengths $r_0$ for samples of galaxies and
  clusters of various depth. Open circles - galaxies in the direction
  of the Coma supercluster, crosses - galaxies in the direction of
  Perseus supercluster, triangles - rich clusters of galaxies.
  (\citet{Einasto:1986oh}).}
\label{KlypinFig}
\end{figure}

\citet{Einasto:1986oh} derived the spatial 3D CF for a series of
galaxy and cluster samples of different depth and luminosity limits in
directions to the Coma and Perseus superclusters. This study
demonstrated that the correlation length depends not only on galaxy
luminosity but also on the depth of the sample, as illustrated in
Fig.~\ref{KlypinFig}. Luciano Pietronero \citep{Pietronero:1987aa}
interpreted this dependence of correlation length on sample depth as
evidence for the fractal nature of the galaxy distribution. A more
detailed analysis of the fractal behaviour of galaxies was carried out
by \citet{Jones:1988nu}. Fractal properties of the galaxy distribution
were discussed at the IAU Symposium “Large Scale Structures of
the Universe,” held in Balatonfured, Hungary, on June 15–20, 1987,
where Bernard Jones presented the main results of
\citet{Jones:1988nu}.

A well-known dialogue between the Anglo-American and Italian schools
of thought on fractal properties of the Universe took place during the
250th anniversary celebration of Princeton University
\citep{Turok:1997aa}. Marc \citet{Davis:1997aa} presented the
Anglo-American view, while Luciano \citet{Pietronero:1997aa} outlined
the Italian perspective. Davis and Pietronero even made a friendly
bet—judged by Neil Turok—over a case of fine Italian or Californian
wine. Davis argued that the correlation length of volume-limited
samples does not increase considerably with sample depth and that
fractal behaviour is limited to scales $0.01 \le r \le
10$~\Mpc. Pietronero, in contrast, claimed that the correlation length
increases with sample depth and that fractal behaviour continues to
arbitrarily large scales.

These contrasting interpretations led to very different pictures of the
fractal characteristics of the cosmic web. It became clear that a new,
independent analysis was needed, based on modern observational data and
state-of-the-art simulations. Such a study was carried out by
\citet{Einasto:2020aa, Einasto:2021ti}. The analysis proceeded in two
steps: first, various statistical methods for studying fractal
behaviour were compared; second, CFs and their
derivatives were analysed for 3D and 2D samples of both model and
observed galaxy distributions.

\section{Statistics of galaxy clustering  \label{spatial}}

To quantify the distribution of galaxies, the two-point correlation
function   was introduced by \citet{Peebles:1973a}. As discussed by
\citet{Peebles:1980aa}, the angular CF of almost all galaxy samples is
well described by the power-law form given in Eq.~(\ref{xilaw}).

The natural estimator for determining the two-point correlation
function is
\begin{equation}
  \xi_N(r) = {DD(r) \over RR(r)} - 1,
  \label{nat}
\end{equation}
where $r$ is the separation between galaxy pairs, and $DD(r)$ and
$RR(r)$ are the normalized counts of galaxy–galaxy and random–random
pairs at distance $r$. Normalization ensures that the total number of
$DD(r)$ pairs equals the total number of $RR(r)$ pairs. Because
galaxies are clustered, the number density of galaxies is enhanced at
small separations, so $DD(r) > RR(r)$ and $\xi(r) > 0$. At large
distances, galaxies are less numerous than the mean density (many
galaxies reside in clusters), so $DD(r) < RR(r)$ and, by construction,
$\xi(r) < 0$, see the inset of Fig.~\ref{MaddoxFig}.

\begin{figure}[h]
\centering 
\resizebox{0.98\textwidth}{!}{\includegraphics*{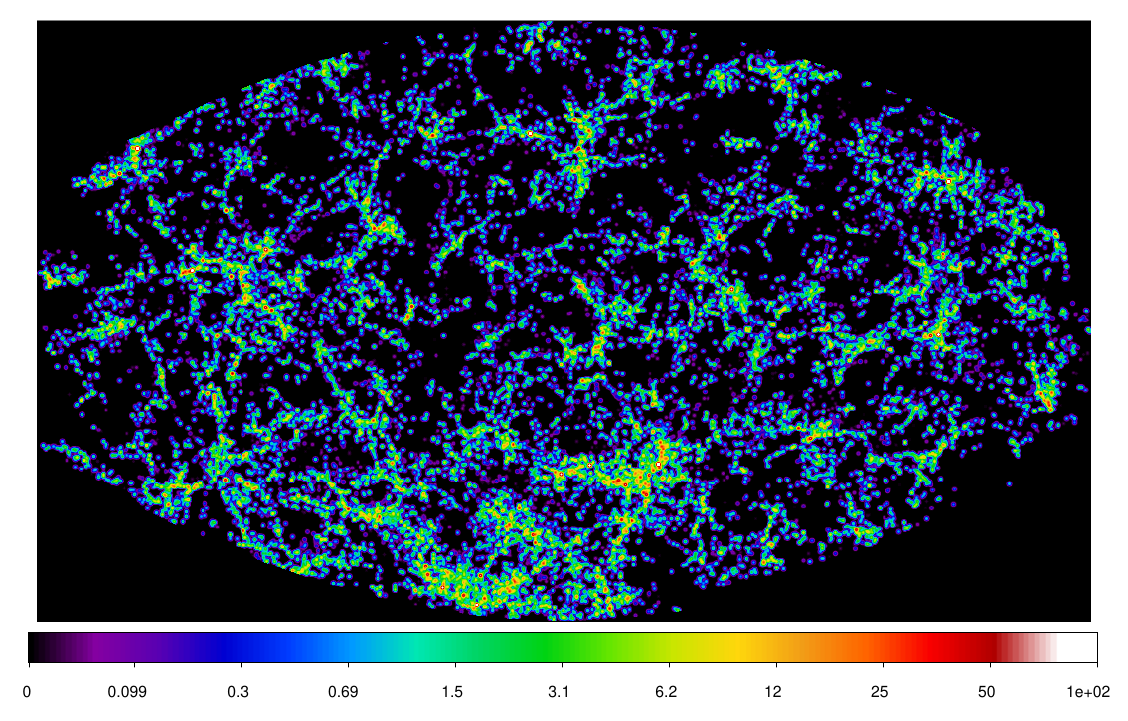}}
\caption{Slice of the density field from the Sloan Digital Sky Survey
at a distance of $240$~\Mpc\ and thickness of $10$~\Mpc.
The Sloan Great Wall is visible in the lower part of the
figure. Credit: \citet{Suhhonenko:2011}, reproduced
with permission © ESO. }
\label{SDSSslice}  
\end{figure}

\citet{Pietronero:1987aa} interpreted the increase of the correlation
length with sample size as a consequence of this normalization effect.
He suggested replacing $\xi(r)$ with an alternative clustering measure,
the structure function $g(r)=1+\xi(r)$, where
$4\pi\,r^2 g(r) n\,\dd r$ gives the mean number of galaxies in a shell
of thickness $\dd r$ at distance $r$ from any point. A similar proposal
was made by Enn Saar \citep{Saar:1989aa}. For a Poisson
distribution, $g(r)=1$. The structure function follows a power law on
small scales ($r < 5$~\Mpc) and approaches zero at large separations.
In the following analysis I use $g(r)$ to investigate the fractal
properties of the galaxy distribution.

\section{Correlation analysis of the cosmic web  \label{cf}}

Historically, quantitative studies of the cosmic web have relied
primarily on CFs. It is well known that the CF captures information
about the amplitudes of the density field but not about its phases
\citep{Coles:2009aa}. This limitation becomes evident when comparing
the left and right panels of Fig.~\ref{fig:Lick} and the SDSS slice
shown in Fig.~\ref{SDSSslice}. Although these distributions have
similar CFs, their visual patterns differ markedly. To quantify the
pattern of galaxy distribution, additional 
statistics are required, such as percolation analysis, first applied
by \citet{Zeldovich:1982}.

\begin{figure}[h]
\centering 
\resizebox{0.48\textwidth}{!}{\includegraphics*{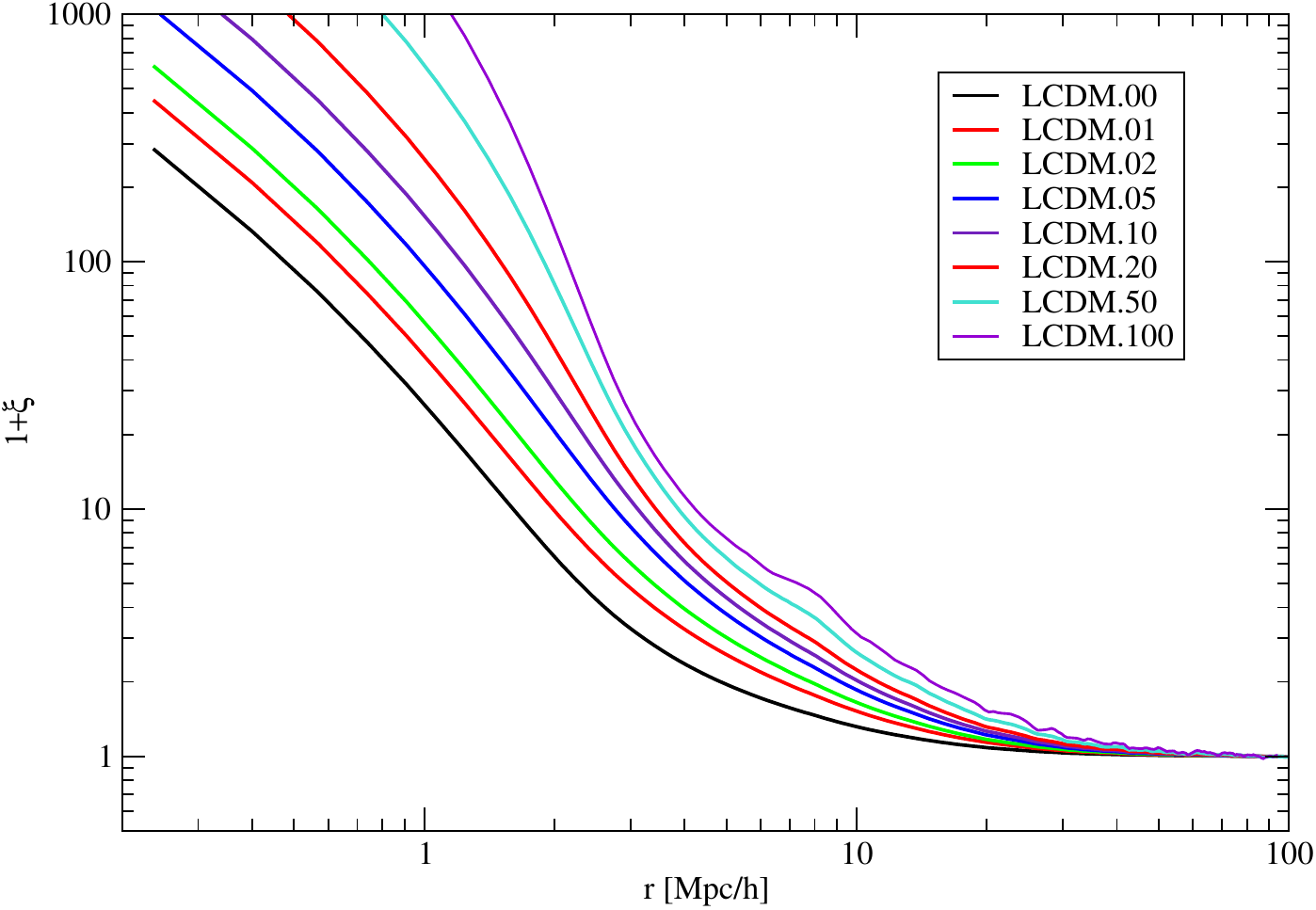}}
\resizebox{0.48\textwidth}{!}{\includegraphics*{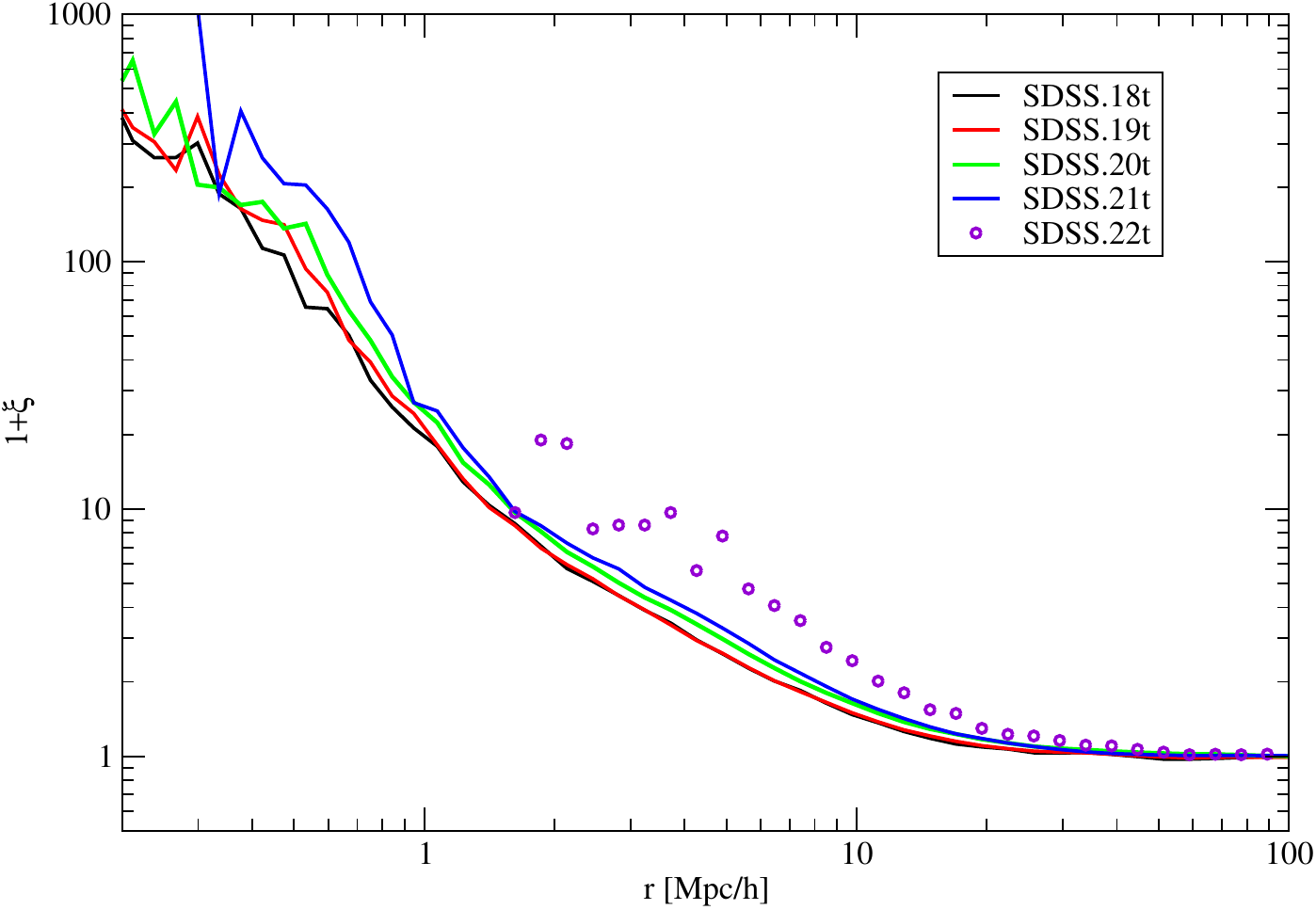}}
\caption{Structure functions defined as $g(r)=1+\xi(r)$. Left panel:
  $\Lambda$CDM model with box size 512~\Mpc\ for different particle
  selection limits. Right panel: SDSS galaxies using five luminosity
  thresholds. Credit: \citet{Einasto:2020aa}, reproduced
with permission © ESO. }
\label{fig:corrFig2} 
\end{figure}

There exist various methods to study the fractal character of samples
of data points, for an overview see \citet{Martinez:2002}.  In the
present study, I use the simplest method, based on the CF and its
derivatives. These include the structure function $g(r)=1+\xi(r)$ and
its logarithmic gradient,
\begin{equation}
  \gamma(r)= {d \log g(r) \over d \log r},
  \label{gamma}
\end{equation}
which I refer to as the $\gamma(r)$ function. This function defines the
fractal dimension,
\begin{equation}
 D(r) =3+ \gamma(r).
  \label{dim}
\end{equation}

\begin{figure}[h]
\centering 
\resizebox{0.48\textwidth}{!}{\includegraphics*{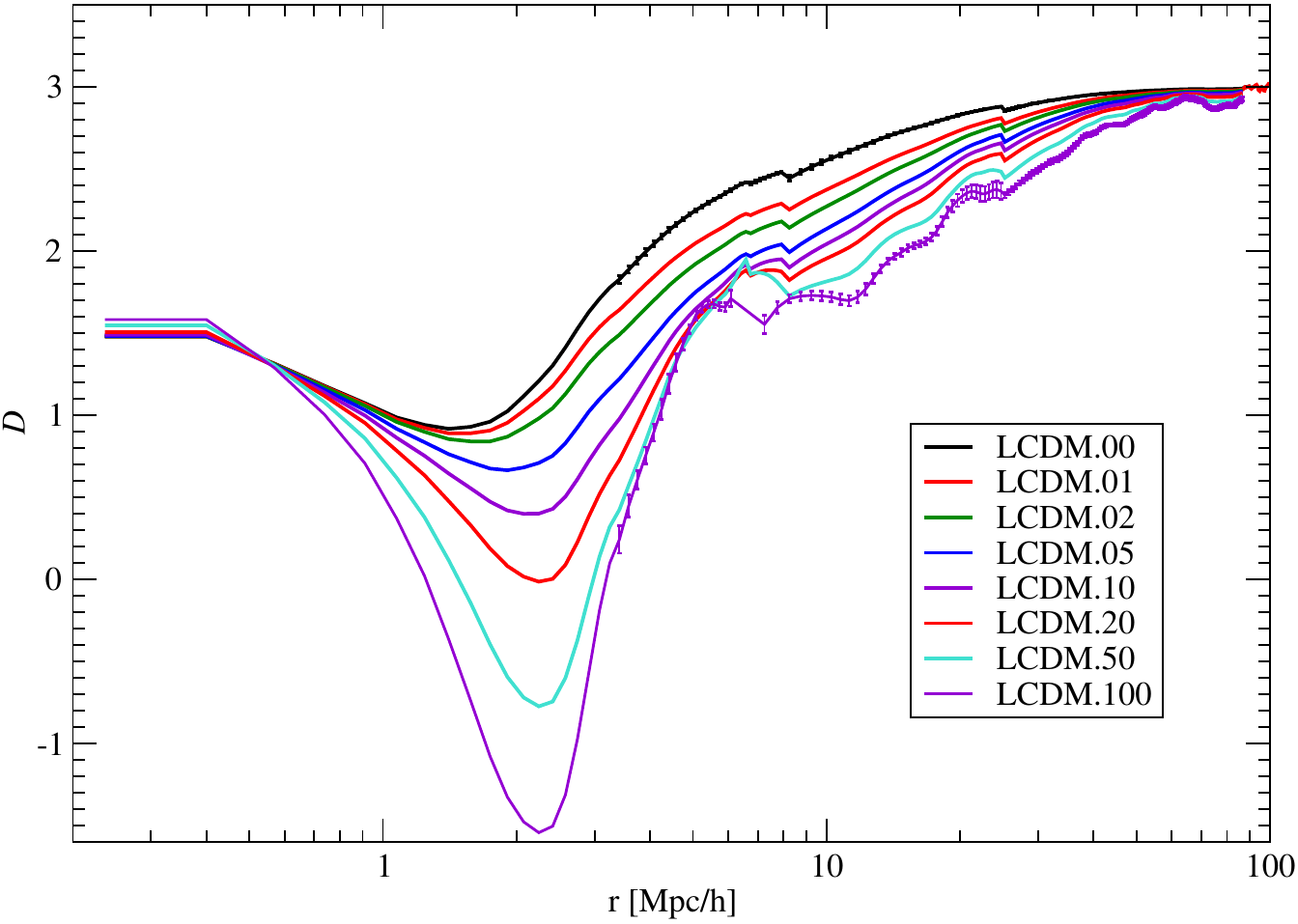}}
\resizebox{0.48\textwidth}{!}{\includegraphics*{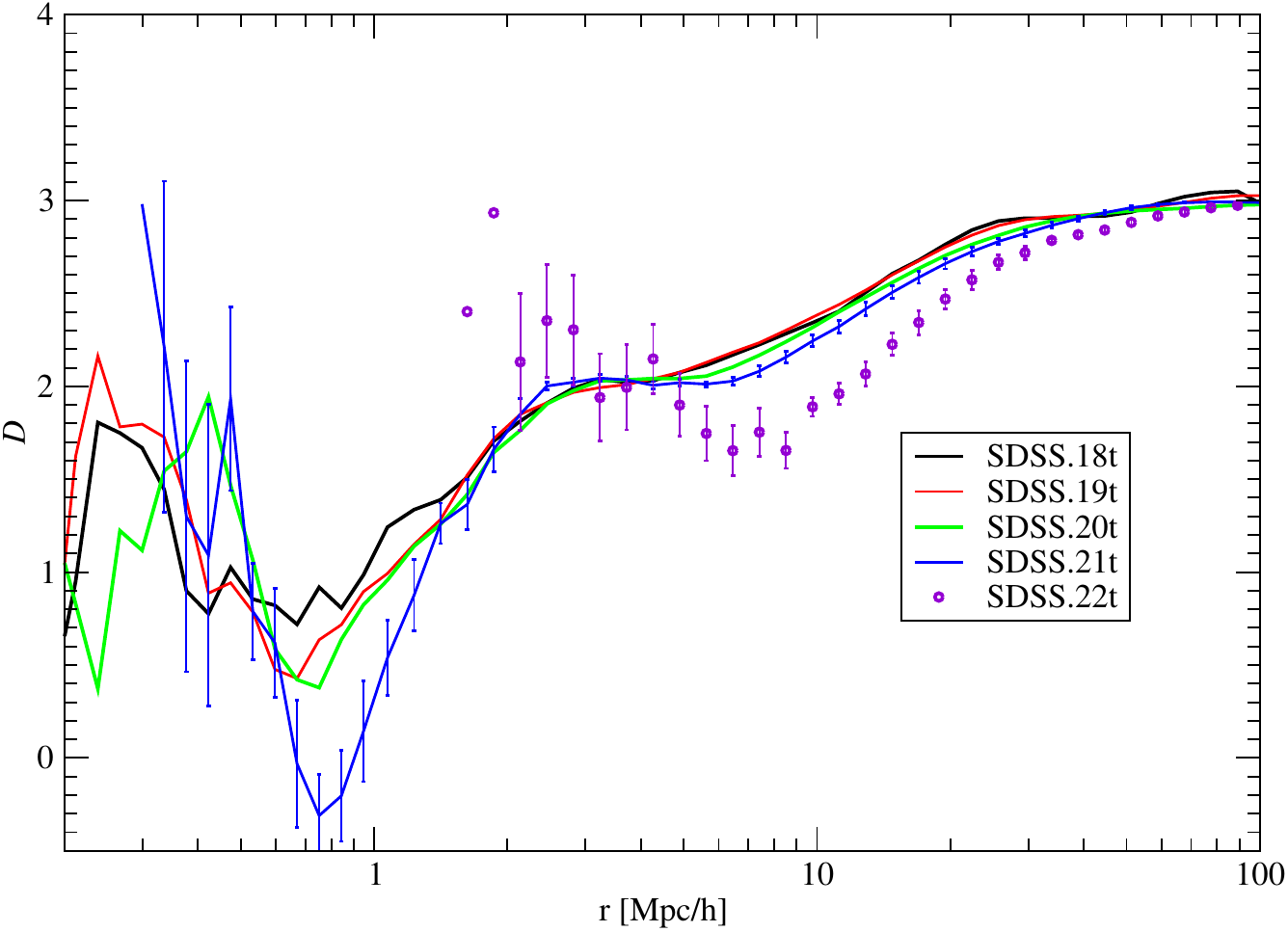}}
\caption{Fractal dimension functions expressed as $D(r)=3+\gamma(r)$.
  Panel arrangement corresponds to Fig.~\ref{fig:corrFig2}. 
  Credit: \citet{Einasto:2020aa}, reproduced
with permission © ESO.}
\label{fig:corrFig3} 
\end{figure}

\citet{Einasto:2020aa} computed correlation, structure, and gamma
functions for $\Lambda$CDM simulations with box size $512$~\Mpc\ and
for an SDSS sample of comparable volume. The resulting structure
functions are shown in Fig.~\ref{fig:corrFig2}, and the corresponding
fractal dimension functions in Fig.~\ref{fig:corrFig3}. The functions
were calculated for various particle density limits in the simulations
and luminosity thresholds in the SDSS data.

These figures show that both functions reveal two key properties of the
cosmic web: (i) their dependence on particle density or luminosity
limits, and (ii) the presence of two characteristic scale regions. The
amplitudes of the structure functions are higher for more luminous
galaxies, reflecting the well-known bias effect described by
\citet{Kaiser:1984}. The fractal dimension function exhibits two
distinct regimes, with a transition near $r \approx 3$~\Mpc. On small
scales, the fractal dimension characterizes the internal structure of
halos (clusters), while on larger scales it reflects the distribution
of halos and clusters along the filaments of the cosmic web. This
difference in fractal behaviour was already noted in early 3D CF
calculations by \citet{Zeldovich:1982} and \citet{Zehavi:2004aa}. The
transition scale corresponds to the typical diameter of halos/clusters.
At the largest separations, the fractal dimension approaches
$D(100)=3.0$.

\section{Comparing angular and spatial distributions of
  galaxies  \label{2d3d}} 

Redshifts are affected by the internal motions of galaxies within
clusters, producing the Finger-of-God effect.  Galaxies and clusters
experience also coherent motions toward gravitational attractors --
the \citet{Kaiser:1987aa} effect.  To avoid the Kaiser effect,
\citet{Davis:1983ly} recommended using galaxy positions and velocities
independently, i.e., estimating CFs from 2D angular data. This
approach assumes that the 2D and 3D distributions of galaxies are
statistically equivalent. To test this assumption,
\citet{Einasto:2021ti} calculated CFs and their derivatives for a
series of $\Lambda$CDM and Millennium \citep{Springel:2005} models.

\begin{figure}[h]
  \centering 
\resizebox{0.48\textwidth}{!}{\includegraphics*{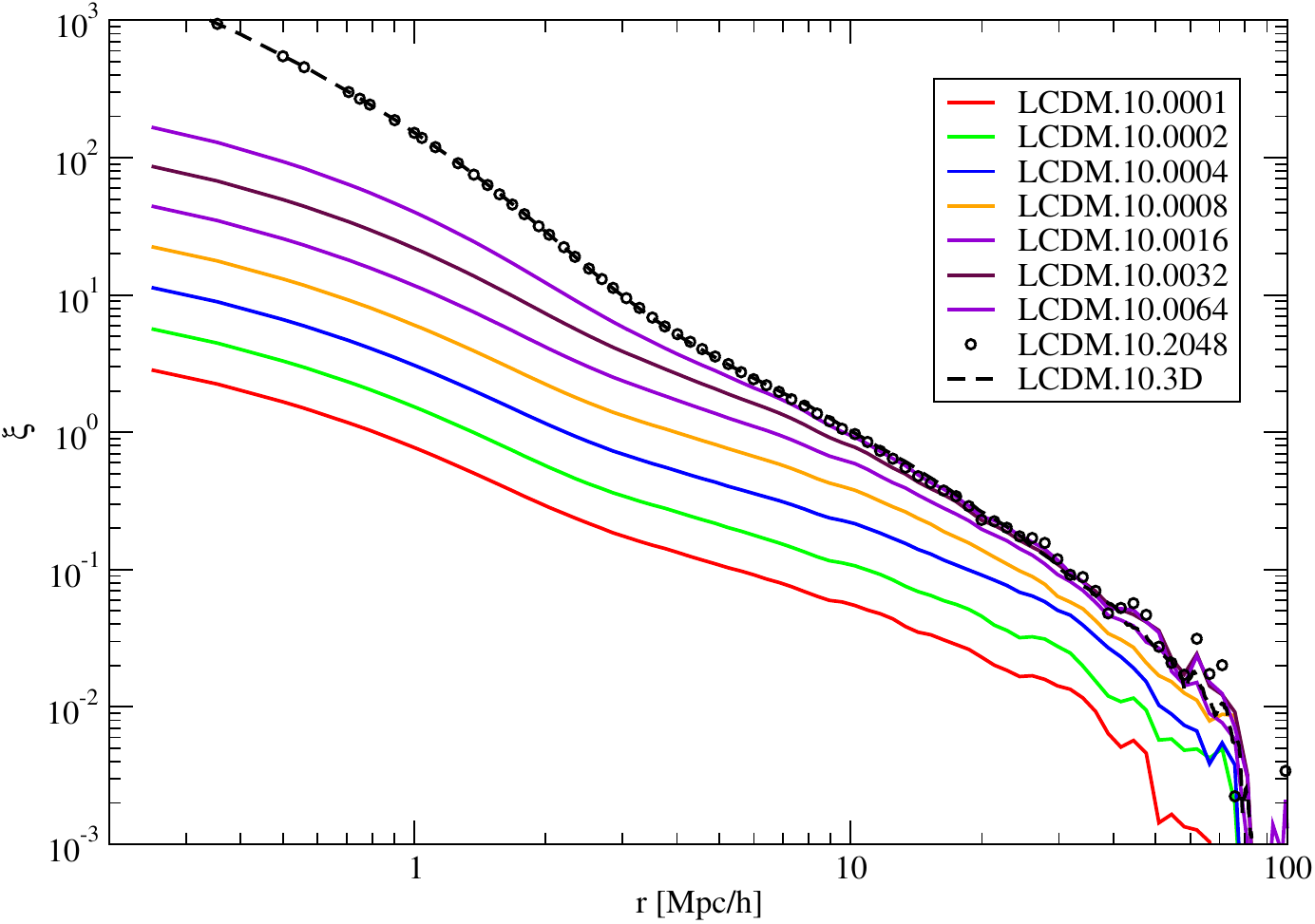}}
\resizebox{0.48\textwidth}{!}{\includegraphics*{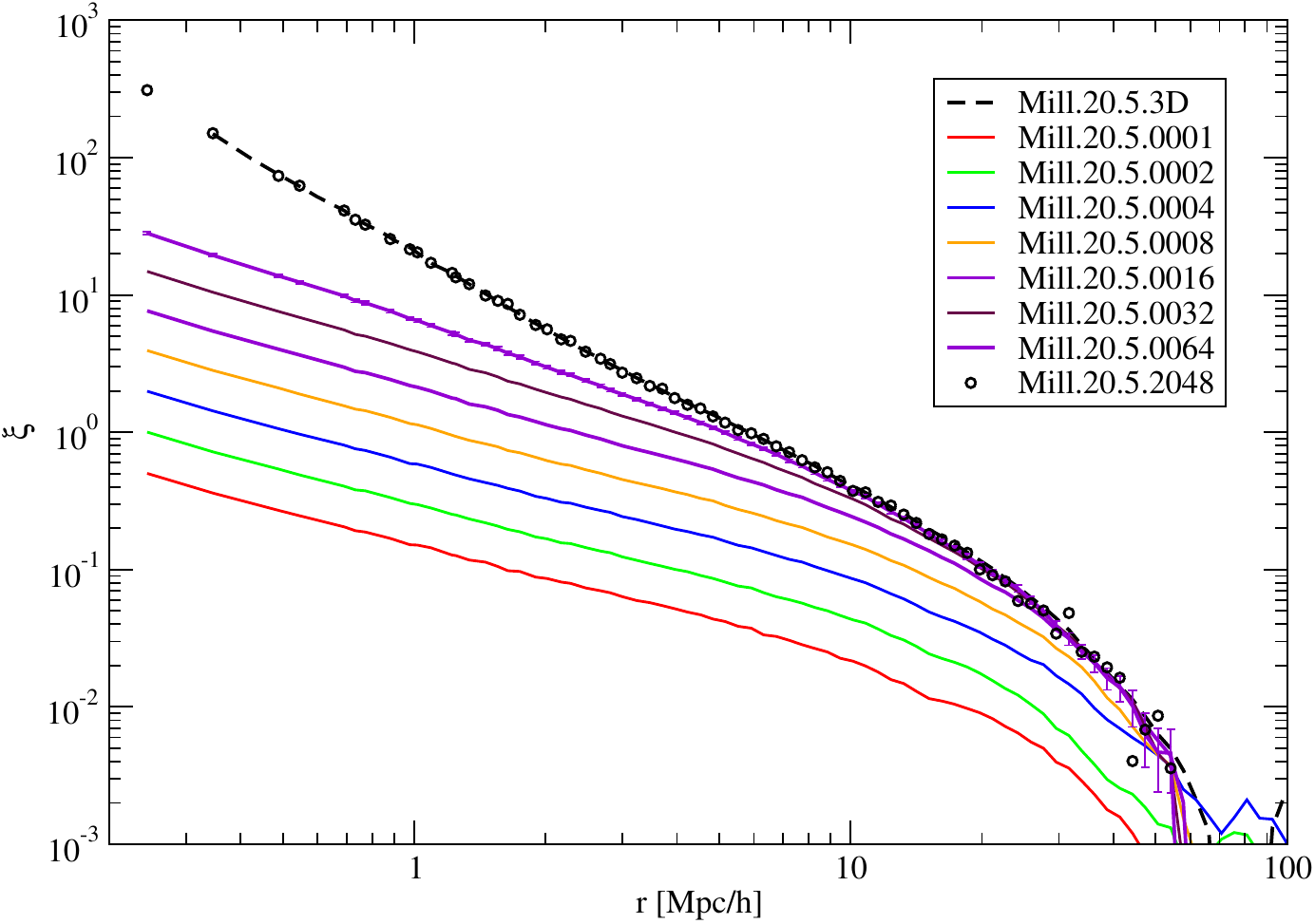}}
\caption{Left: Two-dimensional CFs of the $\Lambda$CDM model with a
  particle density threshold of $\rho_0=10$, shown for various
  thicknesses of 2D samples. Right: Two-dimensional CFs of Millennium
  samples with magnitude limit $M_r=-20.5$, displayed in real space.
  Different colours indicate different sample thicknesses. Dotted lines
  show the corresponding 3D CFs for the same density and luminosity
  limits. Credit: \citet{Einasto:2021ti}, reproduced
with permission © ESO. }
\label{fig:Fig6A} 
\end{figure} 

To estimate 2D CFs, \citet{Einasto:2021ti} divided each 3D sample into
a series of 2D sheets of size $L_0 \times L_0 \times L$~\Mpc, where
$L = L_0/n$ is the thickness of each sheet and $n$ takes values
1, 2, 4, ..., up to 2048. For each $n$, the CFs were computed for all
sheets and averaged to obtain the representative 2D CF for that
thickness.

Figure \ref{fig:Fig6A} shows these CFs for a fixed particle density
limit $\rho_0=10$ in the $\Lambda$CDM model and for Millennium samples
with luminosity threshold $M_r=-20.5$, corresponding roughly to
$L^\ast$ galaxies. The dependence of the 2D CFs on sample thickness $L$
is clearly visible. The case $n=1$ corresponds to the full sample
thickness $L=L_0$ and yields the lowest  amplitude. The thinnest sheets,
with $L=0.25$~\Mpc\ ($n=2048$), produce 2D CFs that closely match the
3D CFs shown by dotted lines. 

According to \citet{Norberg:2001aa} and \citet{Zehavi:2005aa}, the
correlation lengths for the faintest galaxies are
$r_0\approx 4.5$~\Mpc.  As seen in Fig.~\ref{fig:Fig6A}, the
amplitudes of 2D CFs for various thicknesses $L$ are {\em{lower}} than
those of the 3D CFs.  If interpreted as 3D CFs, this underestimates
their correlation lengths -- actual 3D correlation lengths are larger,
depending on the thickness $L$ of 2D samples.
.

\begin{figure}[h]
  \centering 
\resizebox{0.48\textwidth}{!}{\includegraphics*{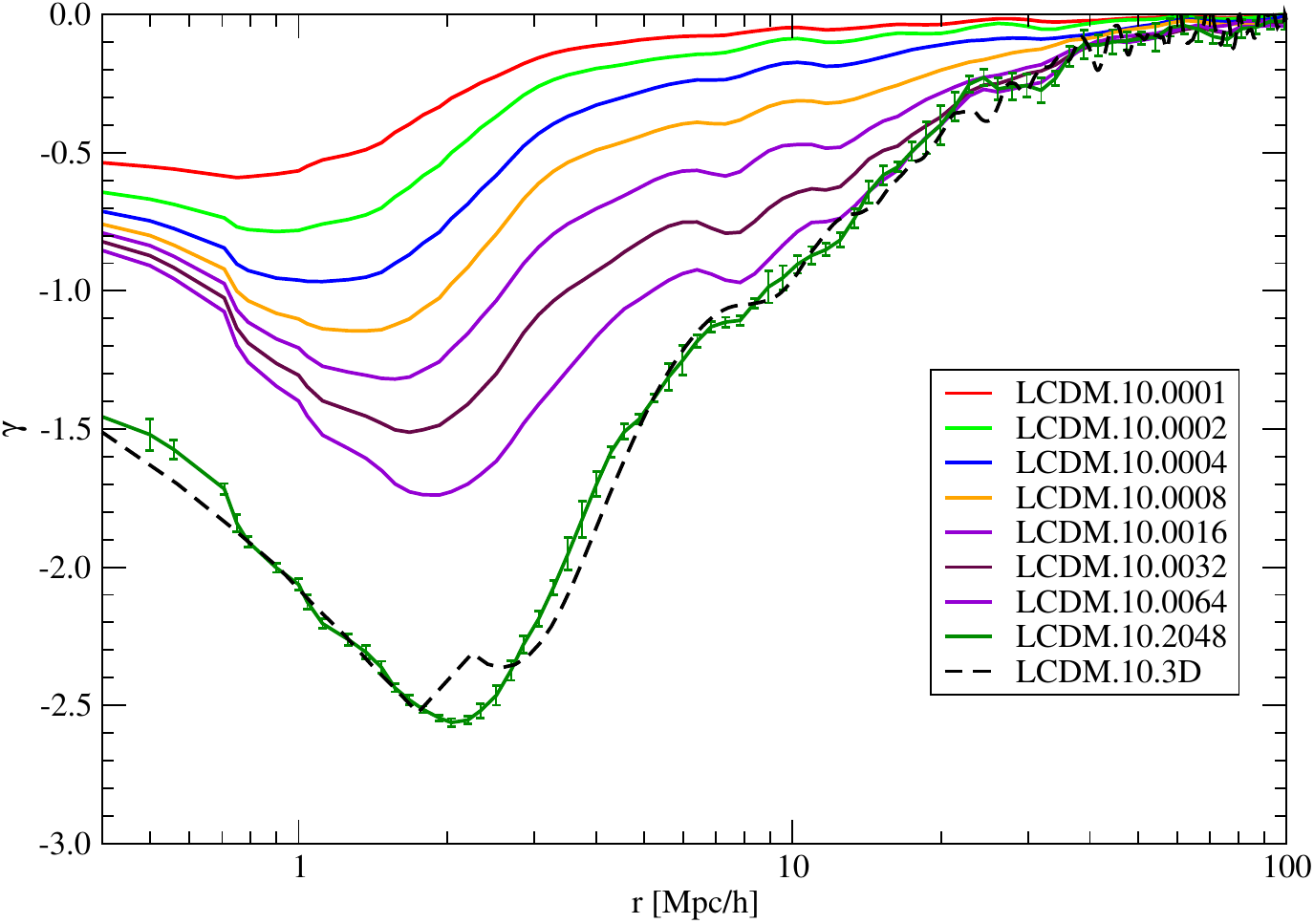}}
\resizebox{0.48\textwidth}{!}{\includegraphics*{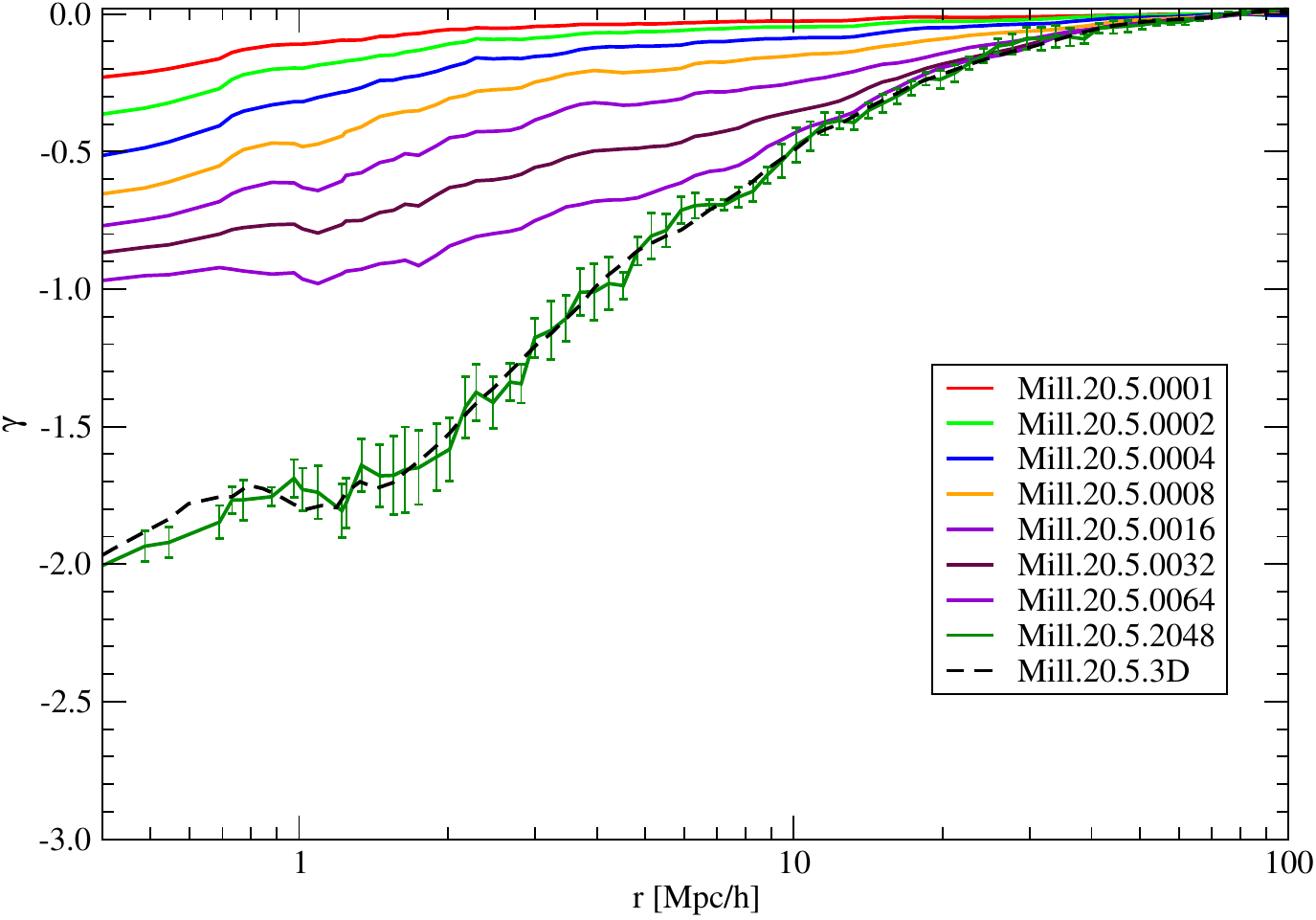}}
\caption{Left: Two-dimensional gradient functions of the $\Lambda$CDM
  model with particle density threshold $\rho_0=10$, shown for various
  sample thicknesses. Right: Two-dimensional gradient functions of
  Millennium samples with magnitude limit $M_r=-20.5$, displayed in
  real space.  Credit: \citet{Einasto:2021ti}, reproduced
with permission © ESO.  }
\label{fig:Fig6B} 
\end{figure} 

Figure \ref{fig:Fig6B} presents the gradient functions for the
$\Lambda$CDM model with density limit $\rho_0=10$ and for Millennium
samples with luminosity threshold $M_r=-20.5$. Curves of different
colour correspond to different sample thicknesses $L$. The fine
structure of halos/clusters in the dark matter $\Lambda$CDM model is
preserved even in thin 2D samples. In contrast, the fine structure of
halos/clusters in the Millennium galaxy samples is lost: at the
luminosity threshold $M_r=-20.5$, clusters contain only a small number
of bright galaxies, making their internal structure invisible.

In 2D Millennium samples with thickness $L\approx 200$~\Mpc, the
gradient reaches $\gamma(r) \sim -0.7$ at small separations and
smoothly approaches $\gamma(r)=0$ at $r=100$~\Mpc. This behaviour
explains the shape of CFs, as found by \citet{Groth:1977aa},
\citet{Davis:1983ly}, and \citet{Maddox:1990aa}, shown in Fig.~\ref{MaddoxFig}..

\section{Discussion  \label{scale}}

\citet{Einasto:1986oh} demonstrated that the correlation length depends
not only on galaxy luminosity but also on the depth of the sample, as
shown in Fig.~\ref{KlypinFig}. This result contradicted the prevailing
understanding of galaxy statistics. I first reported this dependence at
the IAU Symposium on Dark Matter in Princeton in 1985. Although the
paper had not yet been published, the results were already clear. After
my talk, Jim Peebles approached me and asked how such an unexpected
result could be explained. My answer was straightforward: it reflects
the influence of voids.

The CF is defined as $1+\xi(r) = {DD(r) \over RR(r)}$, where $DD(r)$ and
$RR(r)$ are the normalized counts of galaxy–galaxy and random–random
pairs at separation $r$. Consider a volume $V_0$ containing galaxies and
systems of galaxies such as supercluster cores. Let $DD_0(r)$ and
$RR_0(r)$ be the corresponding pair counts. Now surround this volume
with empty space, increasing the total volume to $V_1$. The
galaxy–galaxy counts remain unchanged, $DD_1(r)=DD_0(r)$, but the
random–random counts decrease because the random sample is diluted over
the larger volume $V_1$. This increases the amplitude of $1+\xi(r)$ by a
factor proportional to $V_0/V_1$.

Thus the CF does not measure clustering alone; it also measures the
emptiness of the surrounding volume. At small separations the region
around a cluster (such as Virgo) contains only a small void, but with
increasing sample radius the void volume grows until a statistically
representative volume is reached. As a consequence, the classical
correlation length $r_0=4.5$~Mpc (\ref{xilaw}) systematically
underestimates the true 3D correlation scale of the cosmic web. This
void–volume effect also clarifies why both Davis and Pietronero
misinterpreted the behaviour of the correlation length. Davis relied on
2D angular data, where projection suppresses the contribution of voids
and erases the internal structure of halos. As a result, the amplitude
of the CF is underestimated and the apparent correlation length remains
close to the classical value $r_0=4.5$~Mpc, giving the false impression
that fractal behaviour is limited to very small scales. Pietronero, on
the other hand, interpreted the increase of $r_0$ with sample depth as
evidence for an unbounded fractal hierarchy. However, this increase is
a direct consequence of the growing void volume in deeper samples, not
a sign of fractality extending to arbitrarily large scales. Thus the
void structure of the cosmic web explains both the underestimation of
$r_0$ in 2D analyses and the overinterpretation of its growth in 3D
samples.

\citet{Einasto:2023aa} investigated the evolution of the CF and
its derivatives for $\Lambda$CDM models with box sizes
$L_0=256, 512,$ and $1024$~\Mpc. The amplitudes of CFs for simulated
galaxy samples relative to those of dark matter define the bias
function,
\begin{equation}
  b^2(r, \rho_0) = \xi(r,\rho_0)/\xi(r,0),
  \label{bias}  
\end{equation}
where $\rho_0$ is the particle density threshold used to identify
simulated galaxies. Bias functions exhibit a plateau near
$r\approx10$~\Mpc. The ratio of galaxy CFs to dark matter CFs at this
scale defines the bias parameter, $b^2=\xi_{gal}/\xi_{DM}$. For the
present epoch, \citet{Einasto:2023aa} found bias values of
$b=1.21,~1.29,$ and $1.33$ for the L256, L512, and L1024 simulations,
respectively. This modest increase indicates that all three samples are
already close to being representative.

Wavelet analysis by \citet{Einasto:2011fy} shows that the properties of
the large-scale cosmic web—its filaments and voids—depend on the
synchronization of medium- and large-scale density waves, see also
\citet{Coles:2009aa}. When density 
waves of different scales are synchronized, positive amplitudes combine
to form rich galaxy systems, while negative amplitudes reinforce voids
by lowering the mean density.

\citet{Kofman:1988} demonstrated that the skeleton of the
supercluster–void network is established very early in cosmic history,
shortly after inflation. Structures larger than the scale of matter–
radiation equality become frozen as their corresponding waves leave the
horizon. After perturbations re-enter the horizon, their amplitudes
grow until a characteristic epoch corresponding to redshift
$z\sim 0.7$. The analysis by \citet{Suhhonenko:2011} showed that
density perturbations up to scales $\le 128$~\Mpc\ determine the size
of voids and superclusters. Waves of larger wavelength remain outside
the horizon and do not influence the scale of cosmic structures.

Important constraints on cosmological parameters come from the cosmic
microwave background (CMB) radiation at hydrogen recombination
($z\approx 1000$), when the temperature was about 3000$^\circ$~K. As
emphasized by \citet{Sunyaev:2009aa}, the physics of this epoch is
simple and well understood from laboratory experiments. CMB
observations with the Planck satellite \citep{Planck-Collaboration:2020aa}
yield a spatial curvature $\Omega_k = 0.0007\pm 0.0019$. Planck data
also provide precise estimates for matter densities:
baryon density $\Omega_b\,h^2=0.02233\pm 0.00015$, cold dark matter
density $\Omega_c h^2=0.1198\pm 0.0012$, and dark energy density
$\Omega_\Lambda =0.6889\pm 0.0056$. These values agree well with
constraints from Big Bang nucleosynthesis and from dark matter in
galaxy systems, and they rule out the hypothesis that the mean matter
density approaches zero on large scales.

The scale of homogeneity was estimated by \citet{Maddox:1990aa}, who
suggested that the fractal nature of galaxy distribution holds over
the range 10 kpc $\le r \le 10$~\Mpc, with a fractal dimension of
approximately $D\approx 1.3$. This upper limit 10~\Mpc\ was
interpreted as the scale of homogeneity.

\citet{Bagla:2008aa} defined the scale of homogeneity as the scale at
which deviations $\Delta(D)$ of the fractal dimension fall below its
statistical dispersion. Following this definition, \citet{Yadav:2010aa}
computed these parameters for a $\Lambda$CDM
model with box size 1024~\Mpc\ and estimated the homogeneity scale to
be about 260~\Mpc. \citet{Park:2015aa} analysed a quasar sample from the
SDSS survey and concluded that the concept of a fixed homogeneity scale
is not applicable; instead, homogeneity is achieved only asymptotically
as the observational scale increases.

\section{Summary and Outlook \label{summ}}

Our discussion can be summarised in the following points.

\begin{enumerate}

\item{} The CF is normalised to a Poisson distribution and therefore
  forced to have a negative tail. For this reason it is not suitable
  for measuring large-scale homogeneity. To describe the fractal
  character and its dimension, one should use the structure function
  $g(r)=1+\xi(r)$ and its log–log gradient
  $\gamma(r)= d \log g(r) / d \log r$ instead of the CF.

\item{} Fractal properties of the cosmic web are revealed by the fractal
  dimension function shown in Fig.~\ref{fig:corrFig3}. The fractal
  dimension $D(r)=3+\gamma(r)$ of 3D $\Lambda$CDM and SDSS samples
  (with sizes of 512~\Mpc) characterises the internal structure of
  halos at small separations ($r\le 4$~\Mpc) and the distribution of
  particles/galaxies along filaments at larger separations.

\item{} The gradient function derived from 2D data (Fig.~\ref{fig:Fig6B})
  depends strongly on the thickness $L$ of the samples. In 
  Millennium galaxy samples, the internal structure of halos is not
  preserved in 2D projections, leading to the (incorrect) conclusion
  that the fractal dimension is nearly constant over the interval
  $0.01 \le r \le 10$~\Mpc.

\item{} Early 2D studies suggested that the fractal character of galaxy
  distribution extends only to $\sim 10$~\Mpc, beyond which the
  distribution becomes homogeneous. New 3D data indicate that at large
  separations ($r>100$~\Mpc) the fractal dimension approaches the limit
  $D(r)=3$.

\item{} The scale of homogeneity, as inferred from the fractal dimension
  function and the size of the largest superclusters
  (Fig.~\ref{SDSSslice}), is at least 200~\Mpc.

\end{enumerate}

In recent years, considerable attention has been devoted to analysing
the fractal properties of the early Universe.

These results also clarify why both Davis and Pietronero reached
conclusions that were only partially correct. Davis correctly noted that
the galaxy distribution cannot remain fractal on all scales, but his use
of 2D angular data suppressed voids and halo structure, leading to an
underestimate of the true correlation length. Pietronero, in turn,
rightly emphasised the usefulness of the structure function
$g(r)=1+\xi(r)$, but his assumption of a constant fractal dimension
extending to arbitrarily large scales is incompatible with the physics
of structure formation. The behaviour of density waves, the finite
cosmic horizon, and modern cosmological parameters all show that
fractal properties are limited to a finite interval of scales. Thus both
viewpoints captured important aspects of the problem, but neither
provided a complete description of the cosmic web.

\begin{acknowledgments}
Our special thanks are to   colleagues in Tartu Observatory for
discussions.  Figs. 2 and 3 are reproduced by permission of the
Monthly Notices of the Royal Astronomical Society. 
\end{acknowledgments}

\section*{Funding}
This work was supported by  Tartu Observatory, University
  of Tartu.


\end{document}